\documentclass[prb,onecolumn,showpacs,preprintnumbers,amsmath,amssymb]{revtex4}

\usepackage{graphicx}

\begin{document}

\title{Exciton and biexciton energies in bilayer systems}

\author{M.~Y.~J.~Tan}

\affiliation{St.~Edmund's College, Mount Pleasant, Cambridge CB3 0BN,
UK}

\author{N.~D.~Drummond and R.~J.~Needs}

\affiliation{TCM Group, Cavendish Laboratory, University of Cambridge,
Madingley Road, Cambridge, CB3 0HE, UK}

\date{\today}

\begin{abstract}
We report calculations of the energies of excitons and biexcitons in
ideal two-dimensional bilayer systems within the effective-mass
approximation with isotropic electron and hole masses.  The exciton
energies are obtained by a simple numerical integration technique,
while the biexciton energies are obtained from diffusion quantum Monte
Carlo calculations.  The exciton binding energy decays as the inverse
of the separation of the layers, while the binding energy of the
biexciton with respect to dissociation into two separate excitons
decays exponentially.
\end{abstract}





\maketitle

Bound excitons and biexcitons have been observed in semiconductors
under a variety of conditions.  In this paper we consider bilayer
systems in which the electrons are spatially separated from the holes,
leading to what are known as ``indirect excitons.''  Such systems have
been realized in double-quantum-well structures under an applied
perpendicular electric field, which serves to confine electrons in one
well and holes in the other.\cite{experiments_butov,experiments_snoke}
The possibility of Bose-Einstein condensation of excitons in such
structures has recently aroused much
interest~\cite{experiments_butov,experiments_snoke,butov_2004,rapaport_2004}
and there is a need for a deeper understanding of the processes which
laser excitation initiates in these systems.  In this paper we
consider an aspect of excitations in coupled quantum wells which may
be relevant to experiments on coupled quantum wells - the energetics
of biexcitons in bilayer systems.

The effective-mass approximation with isotropic electron and hole
masses gives a simple description of excitons and biexcitons which has
been applied to many systems.  This model is highly idealized, and
effects due to anisotropic masses, non-parabolic bands, and finite
well widths and depths will be significant.  The model is, however,
simple enough to be solved to very high accuracy, providing benchmark
results while still permitting comparisons with experimental data.  We
have calculated exciton and biexciton energies within the
effective-mass approximation for a system consisting of ideal
two-dimensional electron and hole layers separated by a distance $d$.

An exciton in an ideal two-dimensional bilayer geometry is described
by the Schr\"odinger equation
\begin{equation}
\label{eq:h_exciton_1}
\left[- \frac{\hbar^2}{2m_e}\nabla^2_e -
\frac{\hbar^2}{2m_h}\nabla^2_h + \frac{e^2}{4 \pi \epsilon_0 \epsilon}
\frac{1}{\sqrt{|{\bf r}_e-{\bf r}_h|^2 + d^2}} \right] \Phi({\bf
r}_e,{\bf r}_h) = E \Phi({\bf r}_e,{\bf r}_h)\;,
\end{equation}
where $m_e$ and $m_h$ are the electron and hole masses, respectively,
and $\epsilon$ is the static dielectric constant of the material.  In
the following, energies are given in terms of the exciton Rydberg,
$Ry^* = \mu_{eh}e^4/(2 (4 \pi \epsilon_0 \epsilon)^2 \hbar^2)$, and
lengths are given in terms of the exciton Bohr radius, $a_B^* = 4 \pi
\epsilon_0 \epsilon \hbar^2/(\mu_{eh}e^2)$, where $\mu_{eh} =
m_em_h/(m_e+m_h)$ is the reduced mass.

\begin{figure}[ht]
\includegraphics[width=10cm]{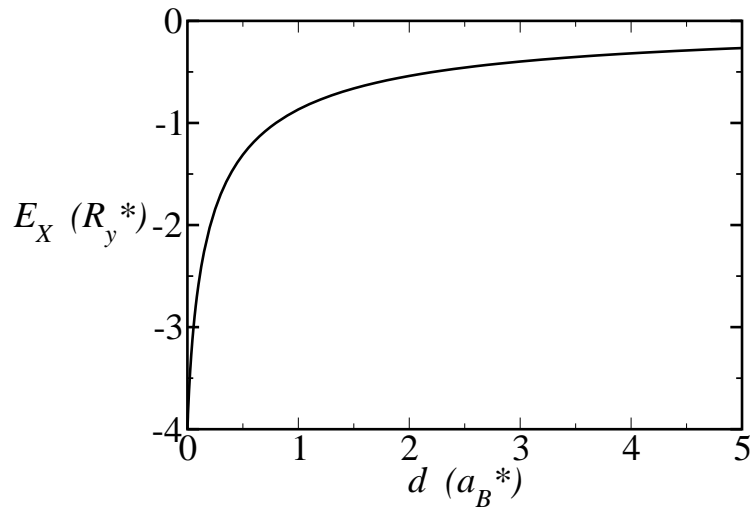}
\caption{Exciton energy as a function of the separation of the
electron and hole layers.
\label{fig:exciton_energy}}
\end{figure}

Eq.~(\ref{eq:h_exciton_1}) may be simplified by transforming into the
center-of-mass frame and separating the variables in cylindrical polar
coordinates.  For the zero-angular-momentum states we obtain
\begin{equation}
\label{eq:h_exciton_2}
- \frac{1}{r}\frac{\partial}{\partial r} \left(r\frac{\partial
  \Phi}{\partial r}\right) - \frac{2}{\sqrt{r^2+d^2}} \Phi = E_X \Phi
  \;,
\end{equation}
where $r$ is the in-plane component of the electron--hole separation.
Eq.~(\ref{eq:h_exciton_2}) may be solved analytically when $d=0$,
which gives a ground state wave function of $\Phi(r) \propto
\exp[-2r]$ and an energy of $E_X = -4\ Ry^*$.  For $d > 0$ we solved
Eq.~(\ref{eq:h_exciton_2}) using a standard Runge-Kutta numerical
integration technique. The exciton energy is plotted as a function of
$d$ in Fig.~\ref{fig:exciton_energy}. The energy takes its minimum
value of $-4\ Ry^*$ at $d=0$, while at small separations the energy
varies linearly with $d$ and at large separations it varies as $1/d$.
The results shown in Fig.~\ref{fig:exciton_energy} may be fitted to
the expression
\begin{equation}
\label{eq:exciton_energy_fit}
E_X = -\frac{4 + A d + B d^2 + Cd^3}{1 + D d + E d^2 + F d^3 +
Gd^4}\;,
\end{equation}
where $A = 154.363$, $B = 648.9$, $C = 225.005$, $D = 46.4263$, $E =
384.976$, $F = 628.158$, and $G = 129.672$.  This expression gives a
maximum error of less than $0.0028\ Ry^*$ in the range $0<d<10\
a_B^*$.

The Schr\"odinger equation for the biexciton is
\begin{equation}
\label{eq:h_biexciton_dimless}
\left[ - \frac{1}{1+\sigma}(\nabla^2_1 + \nabla^2_2) -
\frac{\sigma}{1+\sigma}(\nabla^2_a + \nabla^2_b) + \frac{2}{r_{12}} +
\frac{2}{r_{ab}} - \frac{2}{r_{1a}} - \frac{2}{r_{1b}} -
\frac{2}{r_{2a}} - \frac{2}{r_{2b}}\right] \Psi = E_{XX} \Psi \;,
\end{equation}
where $1$ and $2$ denote the electron coordinates, $a$ and $b$ denote
the hole coordinates, $r_{12} = |{\bf r}_1 - {\bf r}_2|$, $r_{1a} =
\sqrt{|{\bf r}_1-{\bf r}_a|^2 + d^2}$, etc., and $\sigma = m_e/m_h$.
When expressed in units of $Ry^*$, $E_X$ is a function only of $d$,
but the biexciton energy, $E_{XX}$, is a function of both $d$ and
$\sigma$.  Eq.~(\ref{eq:h_biexciton_dimless}) does not separate in
cylindrical polar coordinates and we have to solve the many-body
problem.  For this purpose we have used the diffusion quantum Monte
Carlo (DMC) method, which is a stochastic projector technique for
solving the imaginary-time many-body Schr\"{o}dinger
equation.\cite{foulkes_2001} In the ground state of the biexciton the
electrons have opposite spins and the holes have opposite spins, so
the spatial part of the wave function is node-less.  The DMC method is
exact in principle for node-less wave functions, and although there
are biases due to the use of finite time steps and populations of
walkers, these can be made negligible for small systems such as this.

The sampling within DMC is guided by an approximate wave function
which must be sufficiently accurate to give low statistical noise and
to keep the biases small.  The form of our approximate wave function
was guided by the symmetries of the problem and the long- and
short-distance behavior.  The system composed of two separated bound
excitons is always more stable than one consisting of four unbound
charges.  Therefore we expect the wave function to be exponentially
small when all four particles are far apart.  When one of the
particles is far from the other three we expect the wave function to
be exponentially small because the single charge will be attracted to
the other three.  Likewise we expect that the part of the wave
function corresponding to one bound exciton and a free electron and
hole is exponentially small.  When $d$ is large we expect the system
to consist essentially of two separated excitons, and the form of the
approximate wave function must allow for this possibility.  The short
range behavior of the wave function is fixed by the Kato cusp
conditions~\cite{kato_1957}, which ensure that the divergences in the
potential and kinetic energies cancel when two particles are
coincident.  The biexciton wave function, $\Psi$, should be unaltered
by exchange of (i) the two electron coordinates, or (ii) the two hole
coordinates, i.e., $\Psi({\bf r}_1,{\bf r}_2,{\bf r}_a,{\bf r}_b) =
\Psi({\bf r}_2,{\bf r}_1,{\bf r}_a,{\bf r}_b) = \Psi({\bf r}_1,{\bf
r}_2,{\bf r}_b,{\bf r}_a)$, and when the electron and hole masses are
equal, $\Psi$ should have the additional electron--hole symmetry
$\Psi({\bf r}_1,{\bf r}_2,{\bf r}_a,{\bf r}_b) = \Psi({\bf r}_a,{\bf
r}_b,{\bf r}_1,{\bf r}_2)$.

The binding between the excitons is expected to be small compared with
the binding within an exciton.  We therefore write the wave function
as an appropriately-symmetrized product of two exciton wave functions,
which is then multiplied by a Jastrow function containing
electron--electron and hole--hole terms.  We use the following form,
which satisfies all of the above conditions,
\begin{eqnarray}
\Psi & = & \Psi_{ee} \Psi_{hh} \Psi_{eh} \nonumber \\ \Psi_{ee} & = &
\exp\left[ \frac{c_1r_{12}}{1+c_2r_{12}} \right] \nonumber \\
\Psi_{hh} & = & \exp\left[ \frac{c_3r_{ab}}{1+c_4r_{ab}} \right]
\nonumber \\ \Psi_{eh} & = & \exp\left[ \left (\frac{c_5r_{1a} +
c_6r_{1a}^2} {1+c_7r_{1a}} + \frac{c_5 r_{1b} + c_{8} r_{1b}^2}{1 +
c_{9}r_{1b}} + \frac{c_5r_{2a} + c_{8} r_{2a}^2}{1 + c_{9}r_{2a}} +
\frac{c_5r_{2b} + c_6r_{2b}^2} {1+c_7r_{2b}}\right) \right] \nonumber
\\ & + & \exp\left[ \left (\frac{c_5r_{1a} + c_{8}r_{1a}^2}
{1+c_{9}r_{1a}} + \frac{c_5r_{1b} + c_6 r_{1b}^2}{1 + c_7r_{1b}} +
\frac{c_5r_{2a} + c_6 r_{2a}^2}{1 + c_7r_{2a}} + \frac{c_5r_{2b} +
c_{8}r_{2b}^2} {1+c_{9}r_{2b}}\right) \right] \label{eq:biexciton} \;,
\end{eqnarray}
where $c_1$--$c_9$ are parameters.  This form is similar to that used
by Lee \textit{et al.}~\cite{lee_1983} for the three-dimensional
biexciton, although our form has more flexibility in the
electron--hole part.  We require $c_2,c_4,c_7,c_9 > 0$ so that $\Psi$
is well-behaved, and $c_6, c_8 < 0$ so that $\Psi_{eh}$ decays when
the electrons and holes are far apart.  Eq.~(\ref{eq:biexciton})
describes two separated excitons when either $c_6$ or $c_8$ go to
zero.

The values of the parameters $c_1$ and $c_3$ were fixed by the
electron--electron and hole--hole Kato cusp
conditions.\cite{kato_1957} The value of $c_5$ was fixed by the
electron--hole cusp condition when $d=0$, while for $d>0$ there should
be no electron--hole cusp and so we set $c_5=0$.  When $\sigma=1$
electron--hole symmetry requires that $c_1 = c_3$ and $c_2 = c_4$.
The optimal values of the remaining variable parameters were obtained
by minimizing the variance of the variational
energy.\cite{umrigar_1988,kent_1999}

\begin{figure}[ht]
\includegraphics[width=10cm]{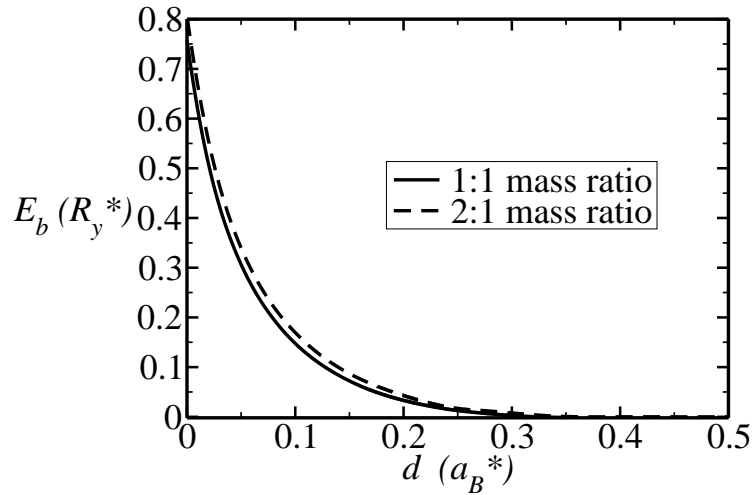}
\caption{The binding energy of the biexciton, $E_b = 2E_X - E_{XX}$,
as a function of the separation of the electron and hole layers, $d$,
for $\sigma=$ 1 and 2.  The error bars are smaller than the thickness
of the lines.
\label{fig:biexciton_energy}}
\end{figure}

We calculated the energy of the biexciton, $E_{XX}$, as a function of
$d$ for $\sigma = $ 1 and 2.  Tests indicated that the timestep and
population control errors in the DMC results were negligible.  All the
QMC calculations were performed using the \textsc{casino}
code.\cite{casino} The biexciton binding energy with respect to
dissociation into two separate excitons, $E_b = 2E_X - E_{XX}$, is
plotted as a function of $d$ in Fig.~\ref{fig:biexciton_energy} for
$\sigma = $ 1 and 2.  For small $d$ and $\sigma = 1$, $E_b$ is close
to the value obtained in earlier calculations for $d=0$ of $0.771\
Ry^*$.\cite{2d_biexciton}
$E_b$ goes to zero at large $d$ much more rapidly than $E_X$ itself
because the electron--electron and hole-hole repulsions dominate the
electron--hole attraction in the biexciton at large $d$, tending to
unbind the biexciton.  Examination of the biexciton wave function
shows that two separated excitons are formed at large $d$.  The
behavior of $E_b$ at large $d$ is reasonably well represented by a
simple exponential form,
\begin{equation}
\label{eq:biexciton_binding_energy_fit}
E_b = \alpha \exp[-\beta d]\;,
\end{equation}
where $\alpha = 0.67573$ and $\beta = 15.023$ for $\sigma = 1$, and
 $\alpha = 0.71714$ and $\beta = 14.187$ for $\sigma = 2$.

As a simple example of the use of these results we estimate the
biexciton binding energy, $E_b$, in the experiments of Butov
\textit{et al.}~\cite{experiments_butov}, who studied a system of two
80 \AA-wide GaAs quantum wells separated by a 40 \AA-wide barrier of
Al$_{0.33}$Ga$_{0.67}$As.  The electron mass in GaAs is $m_e = 0.067
m_0$, while the heavy-hole mass should be reduced from its bulk value
of $0.45 m_0$ by confinement effects, and for simplicity we take a
value of $m_h = 0.134 m_0$.  This gives a 2:1 mass ratio, although the
results are not sensitive to the precise value of $m_h$.  Using a
dielectric constant appropriate to GaAs of 13.2 we find $a_B^* = 156$
\AA\ and $Ry^* = 3.5$ meV.  $E_b$ is sensitive to the value of $d$,
and therefore we fix its value such that we reproduce the large-field
exciton binding energy of 4 meV ($1.14\ Ry^*$) calculated for this
structure by Szymanska and Littlewood~\cite{szymanska_2003}, who used
a realistic description of the finite well widths and depths.  Using
Eq.~(\ref{eq:exciton_energy_fit}) we find $d = 0.64\ a_B^*$ ($100$
\AA), which is a very reasonable value as it lies between the
experimental barrier width of 40 \AA\ and the distance between the
centers of the wells of 120 \AA.\cite{experiments_butov} Substituting
$d=0.64\ a_B^*$ into Eq.~(\ref{eq:biexciton_binding_energy_fit}) and
using the parameters for $\sigma = 2$ we obtain $E_b = 8.2 \times
10^{-5}\ Ry^*$ ($2.9 \times 10^{-4}$ meV).  This model predicts an
extremely small biexciton binding energy, which is unlikely to lead to
measurable effects.

In conclusion, we have calculated the energies of excitons and
biexcitons in ideal two-dimensional bilayer systems within the
effective-mass approximation with isotropic electron and hole masses.
The exciton binding energy decays as the inverse of the layer
separation, while the biexciton binding energy with respect to
dissociation into two separate excitons decays exponentially.  This
model predicts that the biexciton binding energy in the experiments of
Butov \textit{et al.}~\cite{experiments_butov} is extremely small.

\vspace{0.5cm}

We thank Peter Littlewood for useful discussions.  NDD and RJN
acknowledge financial support from the Engineering and Physical
Sciences Research Council (EPSRC), UK.


\begin{thebibliography}{99}


\bibitem{experiments_butov} L.V. Butov, A.C. Gossard, and D.S. Chemla,
Nature (London) \textbf{418}, 751 (2002).

\bibitem{experiments_snoke} D. Snoke, S. Denev, Y. Liu, L. Pfeiffer,
and K. West, Nature (London) \textbf{418}, 754 (2002).


\bibitem{butov_2004} L.V. Butov, L.S. Levitov, A.V. Mintsev,
B.D. Simons, A.C. Gossard, and D.S. Chemla,
Phys. Rev. Lett. \textbf{92}, 117404 (2004).
 
\bibitem{rapaport_2004} R. Rapaport, G. Chen, D. Snoke, S.H. Simon,
L. Pfeiffer, K. West, Y. Liu, and S. Denev,
Phys. Rev. Lett. \textbf{92}, 117405 (2004).

\bibitem{foulkes_2001} W.M.C. Foulkes, L. Mitas, R.J. Needs, and
G.~Rajagopal, Rev. Mod. Phys. \textbf{73}, 33 (2001).

\bibitem{kato_1957} T. Kato, Commun. Pure Appl. Math. {\bf 10}, 151
  (1957).

\bibitem{lee_1983} M.A. Lee, P. Vashishta, and R.K. Kalia,
Phys. Rev. Lett. \textbf{51}, 2422 (1983).

\bibitem{umrigar_1988} C.J. Umrigar, K.G. Wilson, and J.W. Wilkins,
Phys. Rev. Lett. \textbf{60}, 1719 (1988).

\bibitem{kent_1999} P.R.C. Kent, R.J. Needs, and G. Rajagopal,
Phys. Rev. B \textbf{59}, 12344 (1999).

\bibitem{casino} R.J. Needs, M.D. Towler, N.D. Drummond, P.R.C. Kent,
and A. Williamson, \textsc{casino} version 1.6 User Manual, University
of Cambridge, Cambridge (2002).

\bibitem{2d_biexciton} D. Bressanini, M. Mella, and G. Morosi,
Phys. Rev. A \textbf{57}, 4956 (1998); K. Varga, J. Usukara, and
Y. Suzuki, Phys. Rev. Lett. \textbf{80}, 1876 (1998).

\bibitem{szymanska_2003} M.H. Szymanska and P.B. Littlewood,
Phys. Rev. B \textbf{67}, 193305 (2003).

\end{thebibliography}
\end{document}